\documentclass[aip,reprint]{revtex4-1}
\usepackage{graphicx}
\usepackage{epstopdf}
\usepackage{booktabs}
\usepackage{grffile}
\usepackage{amsmath}

\DeclareMathOperator*{\argmin}{arg\,min}

\begin{document}


\title{Calculating Tumor Trajectory and Dose-of-the-day Using Cone-Beam CT Projections} 



\author{Bernard L. Jones}
\altaffiliation{Author to whom correspondence should be addressed: bernard.jones@ucdenver.edu}
\affiliation{Department of Radiation Oncology, University of Colorado School of Medicine}
\author{David Westerly}
\author{Moyed M. Miften}
\affiliation{Department of Radiation Oncology, University of Colorado School of Medicine}


\newcommand{\super}[1]{\ensuremath{^{\textrm{{\tiny #1}}}}}
\newcommand{\subsc}[1]{\ensuremath{_{\textrm{{\tiny #1}}}}}
\newcommand{\enm}[1]{\ensuremath{#1}}
\newcommand{\mat}[1]{\overline{\overline{#1}}}

\begin{abstract}
\textbf{Purpose:} Cone-beam CT (CBCT) projection images provide anatomical data in real-time over several respiratory cycles, forming a comprehensive picture of tumor movement.  We developed and validated a method which uses these projections to determine the trajectory of and dose to highly mobile tumors during each fraction of treatment.

\textbf{Methods}: CBCT images of a respiration phantom were acquired, the trajectory of which mimicked a lung tumor with high amplitude (up to 2.5 cm) and hysteresis. A template-matching algorithm was used to identify the location of a steel BB in each CBCT projection, and a Gaussian probability density function for the absolute BB position was calculated which best fit the observed trajectory of the BB in the imager geometry. Two modifications of the trajectory reconstruction were investigated: first, using respiratory phase information to refine the trajectory estimation (Phase), and second, using the Monte Carlo (MC) method to sample the estimated Gaussian tumor position distribution.  The accuracies of the proposed methods were evaluated by comparing the known and calculated BB trajectories in phantom-simulated clinical scenarios using abdominal tumor volumes.

\textbf{Results}: With all methods, the mean position of the BB was determined with accuracy better than 0.1 mm, and root-mean-square (RMS) trajectory errors averaged 3.8\enm{\pm}1.1\% of the marker amplitude. Dosimetric calculations using Phase methods were more accurate, with mean absolute error less than 0.5\%, and with error less than 1\% in the highest-noise trajectory. MC-based trajectories prevent the over-estimation of dose, but when viewed in an absolute sense, add a small amount of dosimetric error ($<$0.1\%).

\textbf{Conclusions}: Marker trajectory and target dose-of-the-day were accurately calculated using CBCT projections.  This technique provides a method to evaluate highly-mobile tumors using ordinary CBCT data, and could facilitate better strategies to mitigate or compensate for motion during SBRT.

\textit{This manuscript was submitted to Medical Physics}

\end{abstract}


\maketitle 



%
%

\section{Introduction}

As radiotherapy becomes more conformal, the negative effects of tumor motion become more pronounced\super{1}.  Highly targeted techniques, such as Stereotactic Body Radiotherapy (SBRT), rely on the establishment of a very steep dose gradient around the tumor in order to reduce normal tissue effects\super{2}.  Motion blurs the boundary between tumor and normal tissue, making this dose gradient harder to establish.  Interfraction motion can largely be addressed by image-guided therapy techniques such as cone-beam CT (CBCT)\super{3}; however, intrafraction motion presents a more complicated problem.  Specifically, motion makes it more difficult to ensure that the target receives the full prescribed dose, and also adds uncertainty when calculating the dose-of-the-day based on pre-treatment imaging.

A common approach to account for regular intrafraction motion is to use 4-Dimensional CT (4DCT) imaging to characterize the full range of tumor motion, and to treat a target volume which encompasses this range\super{4,5}.  This guarantees full coverage of the tumor but increases dose to normal tissue in the tumor vicinity.  Another approach is to monitor the tumor position (either directly or using surrogates), and to treat only when the tumor lies within some specified region (i.e. “gating”)\super{6,7}.  More recent advances involve modifying or re-directing the treatment beam in real-time in order to compensate for intrafraction motion\super{8,9}.

More accurate methods to account for tumor motion are based on knowing the instantaneous tumor location over time (i.e. the tumor trajectory).  Several methods exist to determine this.  If the tumor is implanted with radiopaque fiducial markers, the position can be triangulated using stereoscopic imaging from two or more view angles\super{10-13}.  Similarly, the tumor can be implanted with electromagnetic transponders\super{7,14}, or can be monitored in the treatment suite using MRI images\super{15}. One major drawback of these methods is that they rely on technologies or treatments not commonly available to the majority of radiotherapy centers, or are only available for use in certain tumor sites.

CBCT is a widely used pre-treatment imaging modality which captures a relatively low-quality 3D image of the patient anatomy in the treatment position\super{3}.  Typical clinical CBCT systems capture roughly 650 projection images, which are used to reconstruct the 3D image. These projections are typically acquired over a 30-120 second timescale, and the reconstructed images exhibit significant blurring due to respiratory motion.  However, the projection images themselves provide a comprehensive picture of the anatomical movement, as they typically observe the motion of several respiratory cycles.  For tumors implanted with fiducial markers, these projections can reveal the motion of the target volume during treatment.  Unfortunately, the clinical utility of these projections is often limited as these data are incomplete with respect to the tumor position, since the process of projection condenses the location of an object in 3D space \textit{(x,y,z)} to only two dimensions \textit{(u,v)} in the imager geometry.  Calculation of the tumor trajectory from these projections would allow for determination of tumor motion and dose-of-the-day estimation for tumors implanted with fiducial markers, which is often the case for tumors of the lung, liver, or pancreas\super{16-18}.  Additionally, this information could be used in the context of Adaptive Radiotherapy\super{19-21}.

The goal of this work was to calculate the tumor trajectory using CBCT projections, and to validate the accuracy of this trajectory in calculating the tumor dose.  Mathematically, determining the instantaneous 3D position of the tumor using the 2D projections acquired during CBCT is an under-defined problem; in each projection we have two measured data points (\textit{u,v} position within the imager geometry) yet wish to calculate three quantities (\textit{x,y,z} position of the tumor).  The projection measurement confines the position to lie along a line between the imaging source and the point of detection, but cannot by itself determine the depth of the position along this line. Methods have been developed to estimate the tumor position using orbiting CBCT projections, mostly in the context of tracking/targeting mobile tumors in real-time\super{22-24}.  These methods choose the tumor location by maximizing the expectation value of tumor position (which maximizes the probability of hitting the tumor).  However, this may not be appropriate for dosimetric evaluation, since the most likely tumor location is generally in the center of the planning target volume (PTV) and may lead to an over-estimation of the tumor dose.  Moreover, tumors which undergo large respiratory motion often exhibit a cyclical trajectory.  Previous methods are built around the assumption that tumor position is distributed randomly about the daily mean location, leading to a Gaussian distribution of random position errors\super{25}.  However, this assumption is broken in the presence of non-random respiratory motion, and a Gaussian distribution of these positions will contain relatively large uncertainty.  For cyclical motion, the true uncertainty is much smaller, since there is a large correlation between position and respiratory phase.

In this work, we extended these methods for more accurate calculation of tumor dosimetry in highly-mobile tumors by incorporating information regarding the respiratory phase.  Using a motion phantom, we validated the accuracy of these methods, and quantified the uncertainty in using CBCT projections to calculate dose-of-the-day in mobile tumors.

\section{Methods}

To quantify the motion of the target, the location of a marker in 3D phantom geometry is calculated by examining the position of that marker in a series of orbiting 2D projection images.  Our method builds on the work of Poulsen \textit{et al}\super{22} and proceeds as follows.  First, we acquire CBCT projections of a phantom in which the position of a small fiducial marker is visible.  Next, the position of the fiducial marker in these projections is determined automatically using a template-matching algorithm.  Then, the acquired images are binned into different phases of the respiratory cycle.  An estimate of the likely distribution of marker positions in each phase is built using a Gaussian approximation of uncertainty\super{22,25}.  Finally, this Gaussian distribution is used to estimate the unresolved component of position, either by calculating the most likely position or by using Monte Carlo methods to sample this distribution. 

\subsection{Definition of geometry}

Let \textit{(x,y,z)} represent an orthogonal basis in 3D space, with the origin corresponding to the treatment and imaging isocenter.  In this work, we consider the patient to lie in the head-first supine position, and choose \textit{(x,y,z)} to correspond to the anterior-posterior (AP) or vertical, left-right (LR) or lateral, and superior-inferior (SI) or longitudinal directions, respectively.  The kV source/imager is oriented orthogonal to the \textit{z}-axis, and orbits the patient in the \textit{x/y} plane (rotation about the \textit{z}-axis).  \enm{\theta} represents the angle of the kV imaging source with respect to the x axis.  Let \textit{(u,v)} describe the axes of the 2D projection image, where u is parallel to the \textit{x/y} plane and \textit{v} is parallel to the \textit{z}-axis.  Note that this coordinate system matches those typically described in the radiotherapy literature (e.g. the seminal work of Feldkamp \textit{et al}\super{26}); however, particular choice of the relative directions of \textit{x, y, z, u, v,} and \enm{\theta} may alter the relationship and/or sign dependence of these quantities.

\subsection{Projection operator}
Let P define the projection of a position \textit{(x,y,z)} into the \textit{(u,v)} plane at an angle \enm{\theta}; in other words, P represents the acquisition of a CBCT projection. SDD is the source-to-detector distance, p is the image pixel size, SAD is the source-to-axis distance, and \enm{o_u,o_v} are the projected locations of the rotational isocenter in the \textit{u,v} coordinates (the so-called “piercing-point” or “principal point”). In Eqs 1-3, \textit{u} and \textit{v} are given in units of pixels in imaging space.

\begin{equation}
(u,v)=P(x,y,z,\theta)
\end{equation}

\begin{equation}
u=\frac{SDD}{p}\frac{x\sin(\theta)+y\cos(\theta)}{SAD-[x\cos(\theta)-y\sin(\theta)]}+o_u
\end{equation}

\begin{equation}
v=\frac{SDD}{p}\frac{z}{SAD-[x\cos(\theta)-y\sin(\theta)]}+o_v
\end{equation}

Consider the following clinical scenario.  A patient is implanted with several radiopaque fiducial markers in order to assist in the delineation and localization of this tumor for treatment.  A CBCT is acquired in order to visualize the patient anatomy on the day of treatment, and involves the acquisition of several hundred images which contain the projection of that anatomy into 2D.  The acquisition of this CBCT scan is represented by performing \textit{P(\enm{\theta})} for values of \enm{\theta} from 0$^{\circ}$ to 360$^{\circ}$, and for each fiducial marker, each image contains a measurement of the marker position \enm{u_\theta} and \enm{v_\theta}.

\subsection{Marker position}
We used the formalism of Poulsen \textit{et al}\super{22} to construct a 3D Gaussian probability density function (PDF) describing the marker position.  A full derivation of this Gaussian formalism is given in Ref 22, and for convenience of the reader, we use the same notation when possible.  This method uses maximum-likelihood optimization to calculate a PDF which best describes the observed motion of a marker in projection images.  Then, using this PDF, the unresolved component of marker position is determined by computing the expectation value of this PDF along the line between the x-ray source and the detection point within the image (Fig 1).

A 3D Gaussian PDF is defined by 9 quantities: mean location \enm{\vec{(r_0 )} = (x_0,y_0,z_0 )}, variance $(\sigma_x,\sigma_y,\sigma_z )$, and covariance $(\sigma_{xy},\sigma_{yz},\sigma_{zx} )$. The variance and covariance together form the covariance matrix $\overline{\overline{A}}$.  From this, the 3D PDF for marker position \enm{\vec{(r )} = (x_0,y_0,z_0 )} was given by:

\begin{equation}
G(x,y,z)=\sqrt{\frac{\det \mat{B}}{(2\pi)^3}}e^{-(\vec{r}-\vec{r_0})^T\overline{\overline{B}}(\vec{r}-\vec{r_0})/2}
\end{equation}

Here, $\overline{\overline{B}}$ denotes the inverse covariance matrix $\overline{\overline{B}}=\overline{\overline{A}}^{-1}$, and $T$ denotes the transpose operator.  The functional form of Eq. 4 along the line from the imaging source to detection point (defined here as $g$) was described by the following:

\begin{equation}
\sigma=\sqrt{\hat{e}^T\overline{\overline{B}}\hat{e}}
\end{equation}
\begin{equation}
\mu=\frac{(\vec{r_0}-\vec{p})^T\overline{\overline{B}}\hat{e}}{\hat{e}^T\overline{\overline{B}}\hat{e}}
\end{equation}
\begin{equation}
K=\sqrt{\frac{\det \mat{B}}{(2\pi)^3}}e^{-(\vec{p}+\hat{e}\mu-\vec{r_0})^T\mat{B}(\vec{p}+\hat{e}\mu-\vec{r_0})/2}
\end{equation}
\begin{equation}
g(t)=Ke^{-(t-mu)^2/(2\sigma)^2}
\end{equation}

Here, $\hat{e}$ describes a unit vector pointing from the observed marker position to the imaging source, and $\vec{p}$ describes the point on the imager at which the marker was detected (converted to patient geometry). 

Using the observed marker positions $(u_i,v_i)$ in each image $i$, optimization was performed to estimate the mean position and covariance matrix which describes the probabilistic marker positions.  By integrating Eq. 8, it can be shown that the probability of finding the marker at a given detector position is linearly proportional to the product $K\cdot\sigma$.  The optimization finds a mean position and covariance matrix which maximizes the total probability of finding the markers in positions observed in each image $i$.  The optimization was performed using the Nelder-Mead simplex search method\super{27}.

\begin{equation}
[\vec{r_0},\mat{B}]=\argmin_{\vec{r_0},\mat{B}} \sum_i -\log (K_i\sigma_i) 
\end{equation}

 \begin{figure*}
 \includegraphics{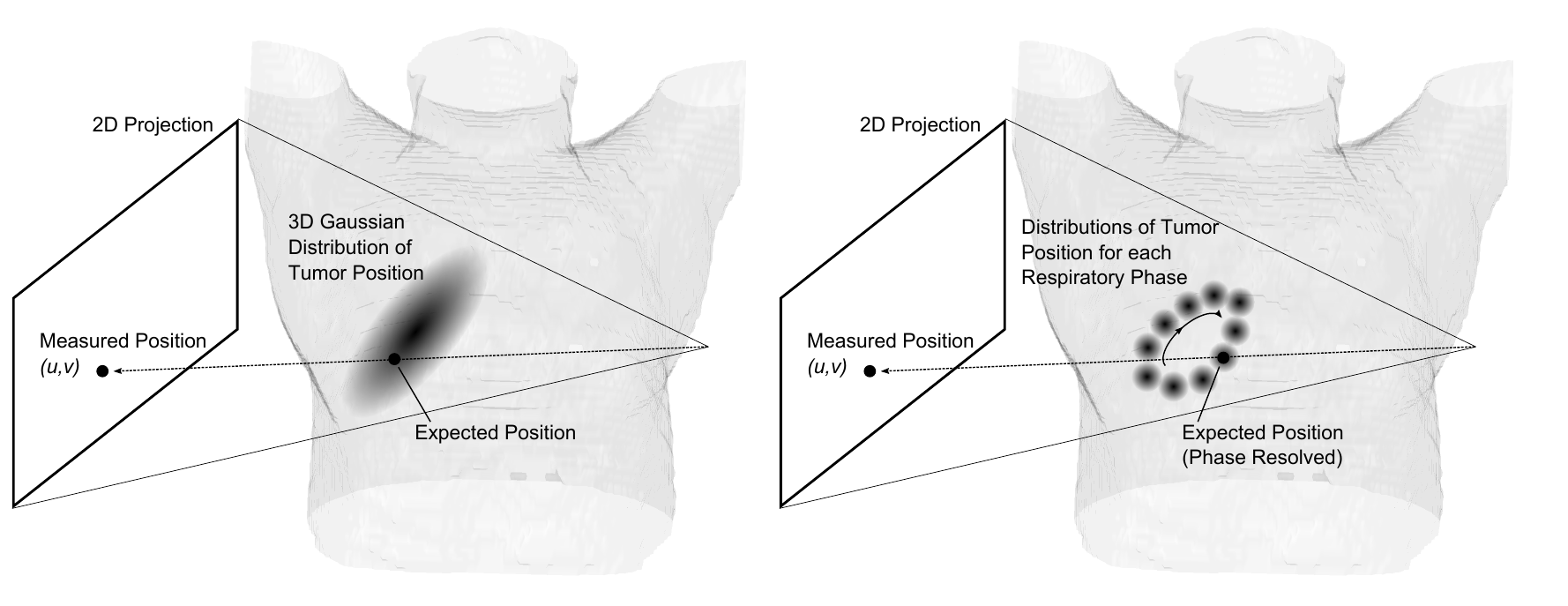}%
 \caption{Determination of tumor position using 3D Gaussian probability.  A) Using all 2D projections, a 3D Gaussian distribution is found which best fits the observed marker trajectory.  For each measured position (i.e. each projection image), the position is determined by finding the expectation of this Gaussian distribution along the line between the x-ray source and the detection point.  B) In the phase-resolved method, a 3D distribution is independently calculated for each phase of the respiratory cycle.}
 \end{figure*}

\subsection{Respiratory phase}

Respiratory phase was determined using the SI displacement of the fiducial markers.  The SI position is well-determined in the CBCT projections for locations near the rotational isocenter, since the only error derives from magnification effects related to $(x,y)$ errors (Eq. 3), which are typically on the order of 1\% or less. This internal surrogate information precluded the need to define respiratory phase based on some other surrogate, such as chest markers or diaphragm movement.

First, the marker trajectory in the $v$ direction was smoothed using a five-sample moving average, and discrete velocity was calculated by taking the difference between smoothed positions in sequential projections. It is assumed that these marker positions represent tumor trajectory in a head-first supine orientation, with the $+v$ axis pointing in the superior direction.  End-exhale $(v_{ee})$ was identified from projections where the marker velocity switched from positive to negative; likewise, end-inhale $(v_{ei})$ was identified where velocity went from negative to positive.

Projections were binned into one of 10 respiratory phases based on the velocity and location of the marker relative to the end-exhale and end-inhale positions.  Phases 1-5 (inhale) consisted of projections with negative velocity, grouped into uniformly spaced bins between subsequent $v_{ee}$ and $v_{ei}$ coordinates.  Phases 6-10 (exhale) consisted of projections with positive velocity, grouped into linearly spaced bins between subsequent $v_{ei}$ and $v_{ee}$ coordinates.  Projections at the beginning (or end) of the motion trace (i.e. those that did not contain measurements of the full cycle) used $v_{ei}$ (or $v_{ee}$) coordinates of the nearest full phase.

\subsection{Position determination}

Four methods were used to determine the 3D marker trajectory from the projection images.  These methods were implemented using in-house software developed with Matlab (The Mathworks, Natick MA). In Method 1 (Base), position estimation followed the method of Poulsen \textit{et al}\super{22}. Data from all projections was used during optimization to calculate a 3D Gaussian distribution which described the probability of finding the marker in a given 3D location (Eq. 9).  This equation was parameterized along the detection ray (the line between the radiation source and the imager detection point), and the marker position along this line was given by $t=\mu$, where $\mu$ equals  the expectation value of the distance along this line (Eqs. 5-8).   

In Method 2 (MC), Monte Carlo methods were used to select the marker position in 3D space in place of the expectation value $\mu$.  First, the detection ray was discretized into 10,000 points, spaced linearly between the source and detector.  The PDF of the 1D Gaussian (Eq. 8) was integrated across the discretized ray to form a normalized cumulative probability distribution function.

\begin{equation}
CDF(t_i)=\frac{\sum_{i=1}^{i=I} g(t_i)}{\sum_{i=1}^{i=i_{max}} g(t_i)}
\end{equation}

Next, a random number $\xi$ was uniformly chosen between 0 and 1. The position of the marker along the ray was chosen to be the point $T$ which satisfied $CDF(T) =\xi$, and this solution was found numerically.

To increase the accuracy of tracking for targets with high respiratory motion, the trajectory of the marker was calculated using a phase-resolved Gaussian approximation (Method 3, Phase).  All projections were binned into one of 10 respiratory phases, and Eq. 9 (optimization of Gaussian distribution) was applied to data from each phase independently.  This resulted in a set of 10 phase-specific mean positions $\vec{r_p}$ and inverse covariance matrices $\mat{B}_p$, or in other words, 10 distinct distributions describing the position of the marker during each phase of respiration (Fig 1B). In Method 4 (PhaseMC), both the Phase and MC methods were applied simultaneously.

 \begin{figure*}
 \includegraphics{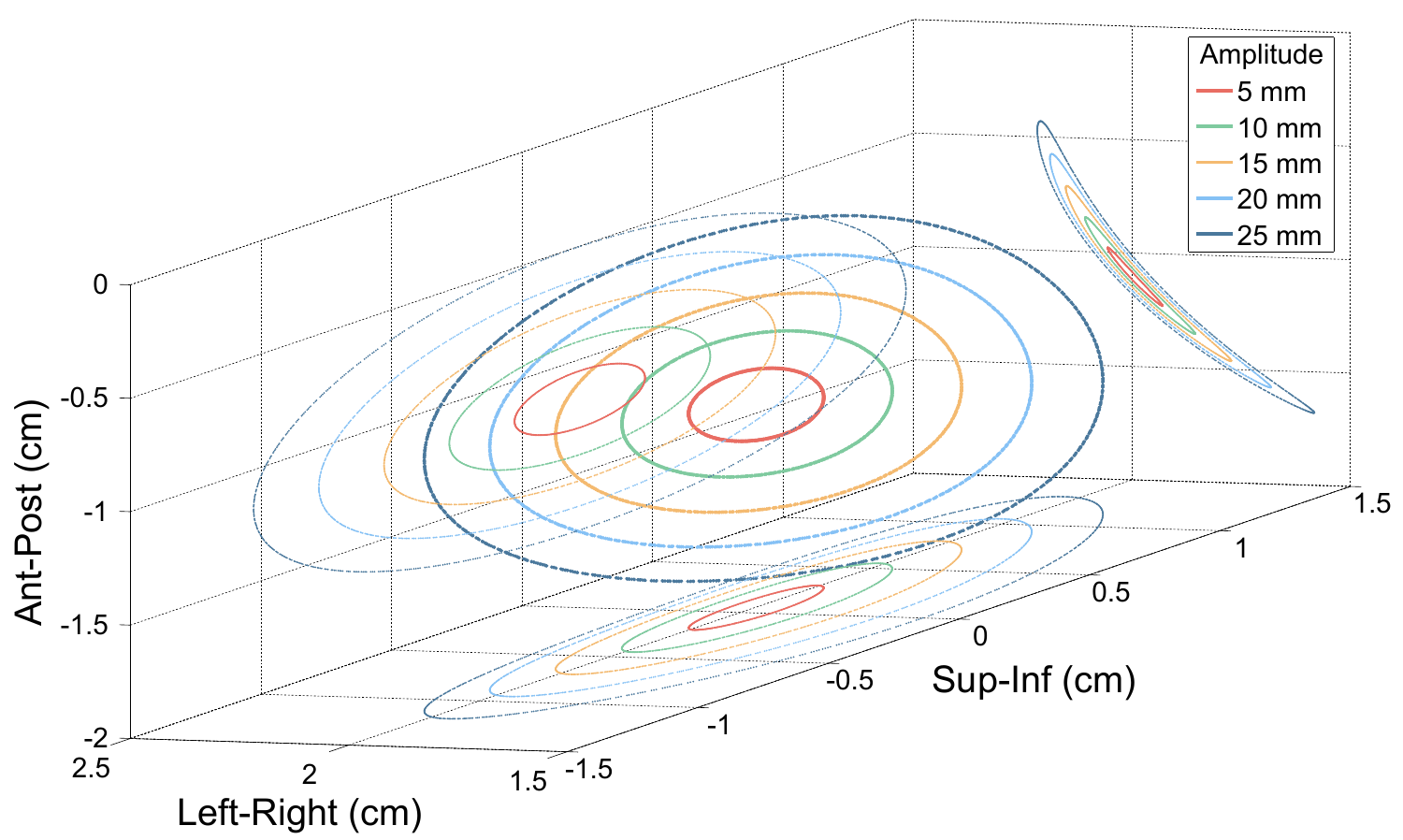}%
 \caption{Trajectory of Motion Phantom.  The motion of the 1-mm-diameter steel BB embedded within the motion phantom is shown relative to the isocenter of the CBCT imaging system.  Trajectories are shown for amplitudes of motion in the Sup-Inf direction ranging from 5 – 25 mm. For clarity, the trajectories are also shown projected onto the XY, XZ, and YZ axes.  Trajectories are shown relative to a standard head-first supine orientation (i.e. the primary axis of motion in the Sup-Inf direction).}
 \end{figure*}

\subsection{Motion phantom}

In order to validate the proposed methods, a phantom was used to simulate respiratory tumor motion (Quasar Phantom, Modus Medical Devices, Inc.).  This phantom contains a 1-mm-diameter steel BB, the trajectory of which is shown in Fig 2 for various superior-inferior (Sup-Inf) amplitudes of motion.  The trajectory is similar to that of a lung tumor trajectory with high amplitude and hysteresis\super{11}.  A CBCT scan of this phantom was acquired using the imaging capabilities of a Varian Truebeam accelerator (Varian Medical Systems, Palo Alto, CA).  656 projection images were acquired using a half-fan trajectory $(\theta=-180^{\circ}\to180^{\circ})$.  The phantom was set to a rate of 11 breaths-per-minute, with a maximum range in the SI direction of 2.5 cm (resulting in maximum AP and LR ranges of 1.2 and 0.6 cm, respectively).  Multiple projection image datasets were acquired with trajectory amplitudes ranging in 1 mm increments from 1 mm to 2.5 cm.  

In each image, the location of the BB was determined automatically using a template matching algorithm\super{28}.  A discretized 2D Gaussian kernel was used as a template image, with width equal to the radius of the BB.  In each image, the normalized cross correlation was computed between the projection image and the template\super{29}. The location of the BB in the image was given by the location where the normalized cross correlation was maximized.  In cases where the template matching algorithm failed, the position was determined manually.  
 
Using the positions of the BB in the projection images, the trajectory of the BB in 3D space was calculated using methods 1-4 (Sect 2.5).  The accuracy of these methods was calculated by comparing the calculated trajectory to the true trajectory (shown in Fig 2).  To simulate more difficult and unpredictable clinical scenarios, additional (simulated) trajectories were created.  Gaussian noise was added to the true marker trajectory by adding normally distributed random values to the $x, y$, and $z$ locations at each point in time. Simulated trajectories were created by varying the standard deviation of this noise from 0 to 1 mm in 0.1 mm increments between datasets.  These altered datasets were projected into 2D image space (Eqs. 1-3), and Methods 1-4 were used to calculate the marker trajectory as described previously.

 \begin{figure*}
 \includegraphics{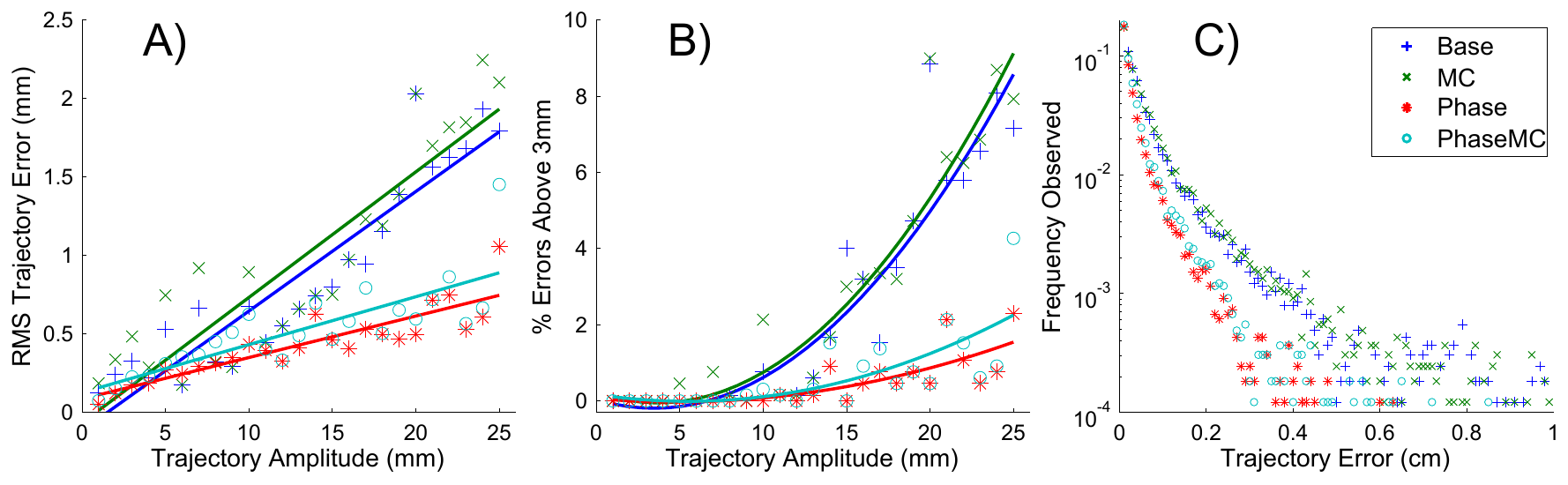}%
 \caption{Errors in calculating trajectory using orbiting views.  Results were calculated with no corrections (Base), with phase corrections (Phase), with Monte Carlo sampling (MC) and with both phase and Monte Carlo (PhaseMC).  Results are shown for phantom motion amplitudes in the SI direction ranging from 1 mm to 25 mm.  A) Root-mean-squared (RMS) errors in the trajectory reconstruction using different methods.  Phase and PhaseMC methods reduce RMS errors by half.  B) Relative fraction of trajectory errors greater than 3 mm.  Phase methods reduce the fraction of large errors, and Monte Carlo methods increase the percentage of large errors slightly.  C) Histogram of trajectory errors in all projections across all acquired datasets.  Phase methods demonstrate increased accuracy.}
 \end{figure*}

\subsection{Dosimetric model}

The goal of this study was to evaluate the accuracy of dosimetric calculations based on reconstructed trajectories.  To that end, the dosimetry of the reconstructed trajectories was compared to the dosimetry of the true trajectory. To evaluate the dosimetry of the true and reconstructed phantom motion trajectories, a phantom-simulated clinical scenario was considered.  A clinical target volume (CTV) was assumed to undergo the motion trajectories under analysis.  Treatment was prescribed to a planning target volume (PTV), which included the CTV plus an isotropic margin.  To simplify the model, and to exaggerate differences between different methods, the dose within the PTV was assumed to equal the prescription, with zero dose outside the PTV.  The PTV (and resulting dose distribution) was stationary over time, while the CTV was translated within the PTV based on either the true phantom trajectory or the reconstructed trajectories.  For each of the 656 points in the trajectory (corresponding to the 656 projection images), the position of the CTV relative to its center was given by the instantaneous trajectory position relative to the mean position.  Based on this position, the dose was calculated for each voxel in the CTV (i.e. based on the binary dose model, or in other words assigning 1/656th of the prescription dose for voxels inside the PTV, and zero for those outside). Independent simulations were performed for CTV-to-PTV margins ranging from 0 mm to 15 mm, and were deliberately set less than the amplitude of motion so as to allow for differences between the proposed methods to become apparent. The goal of our analysis was to quantify the accuracy of dosimetric calculations using CBCT-based trajectory reconstruction.  To that end, our binary dose model serves as a conservative estimate of the inaccuracies of this approach, as any errors in position determination will be reflected in the idealized nature of the target dosimetry simulation.

\subsection{Validation metrics}
We analyzed the accuracy of the proposed methods (Base, MC, Phase, PhaseMC) by comparing these trajectories to the true phantom motion trajectory.  Each CBCT projection dataset (656 images) constituted 656 measurements of the marker position in phantom geometry.  For each measurement, the error was given as the magnitude of the 3D vector difference between the calculated position and the true position.  The root-mean-squared trajectory error was given for each trajectory as the square root of the mean of the squared errors for all measurements in that trajectory.  Additionally, the frequency of large errors was computed by calculating the fraction of measurements with errors exceeding 3 mm.  

The accuracy of dosimetric calculations was computed using our phantom-simulated clinical scenario. Since the dosimetric results depend on the shape of the target, 32 abdominal/lung tumor volumes were used to evaluate the accuracy of the proposed methods (mean volume, 25 $\pm$ 23 cm\super{3}; range 2.0 - 80 cm\super{3}).  The reference dose was calculated by performing this analysis using the exact trajectory, and the calculated dose used the reconstructed trajectories.  The mean dose error was calculated for each volume as the mean dose difference between calculated dose and reference dose.  The mean absolute dose error was calculated as the mean of the magnitude of these dose differences across all voxel of the volume. Mean dose error gives a sense of the dosimetric trends of each trajectory reconstruction method, and can reveal whether a given method underestimates or overestimates dose.  Mean absolute dose error denotes the overall accuracy of a method in determining accurate dose in each voxel (by preventing positive and negative errors from cancelling).  In addition to these metrics, dose-volume histograms of each volume were computed for each method, and compared to the reference. 

 \begin{figure*}
 \includegraphics{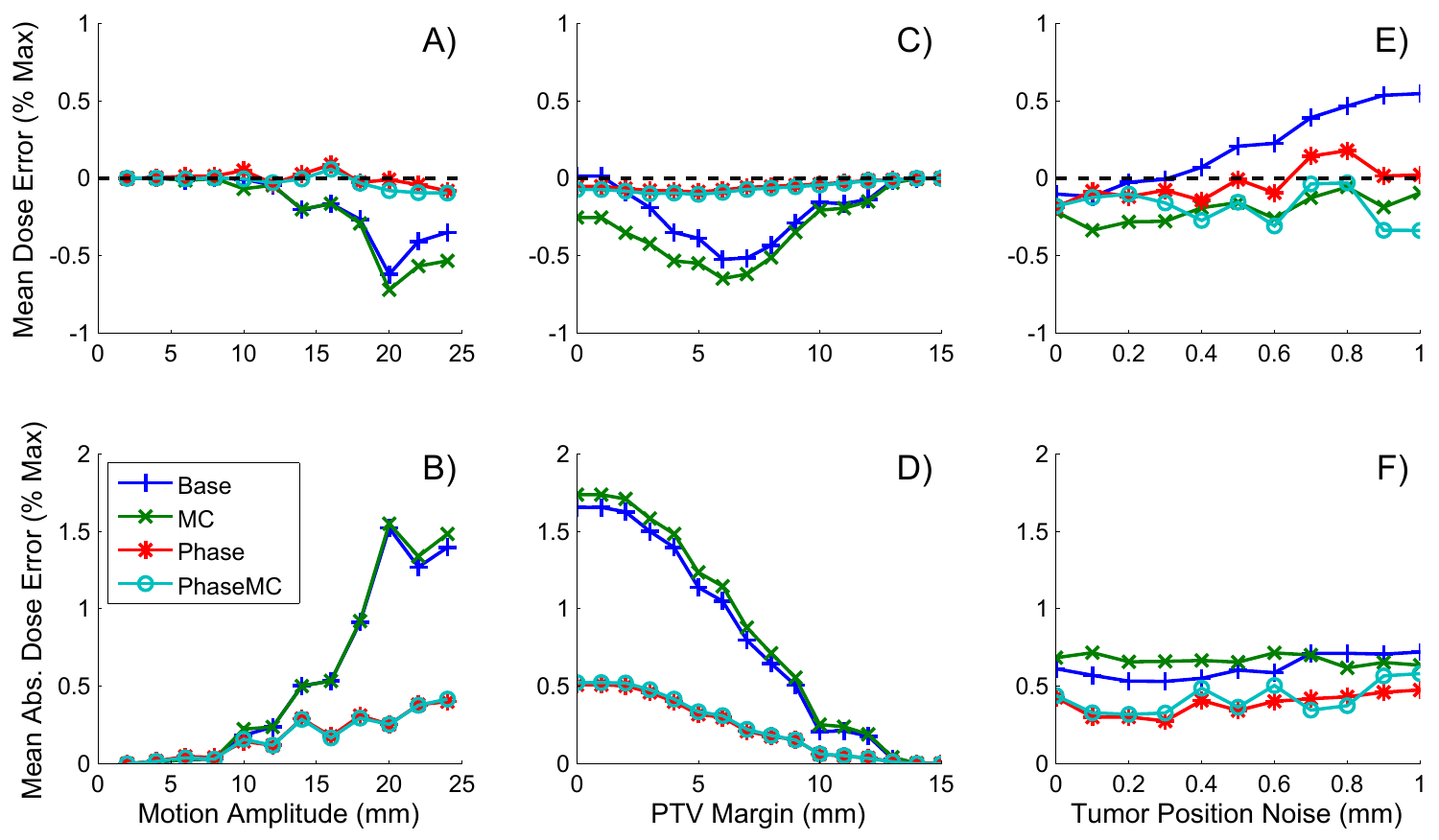}%
 \caption{Mean and mean absolute dose error in the reconstructed trajectories.  Errors were calculated using 32 clinically-drawn abdominal/lung CTVs.  Motion amplitude ranged from 0 to 2.5 cm, and results were calculated with a range of isotropic margins and for trajectories with added noise.  Dosimetric calculations using phase-resolved methods were more accurate, with mean absolute error less than 0.5\% with zero margin, and with error less than 1\% in the high-noise calculation. As noise increases, expectation-based methods generate a trajectory which over-estimates the dose to the CTV.  Monte Carlo-based trajectories under-estimate dose by an average of 0.2\% of prescription dose, even as trajectory noise increases.  In A) and B), the noise is 0 mm and the margin is 4 mm. In C) and D), the amplitude is 2.5 mm and the noise is 0 mm.  In E) and F), the amplitude is 1.2 mm and the margin is 4 mm.}
 \end{figure*}

\section{Results}

\subsection{Accuracy of trajectory reconstruction}

The accuracy of the trajectory calculations are shown in Fig. 3; results were calculated without respiratory phase or Monte Carlo corrections (Base), with phase corrections only (Phase), with Monte Carlo corrections only (MC), and with both phase and Monte Carlo corrections (PhaseMC).  Trajectory amplitude ranged from 0 to 2.5 cm. 
 
With all methods, the mean position of the marker was determined with accuracy better than 0.1 mm.  In Fig. 3a, it can be seen that the Phase and PhaseMC methods reduced the root-mean-squared (RMS) trajectory errors by 50\% compared to the Base and MC methods, with an average RMS error of 3.8 $\pm$ 1.1\% of the trajectory amplitude.  In Fig 3b, the Phase and PhaseMC methods decreased the fraction of large errors (defined here as trajectory error $>$ 3 mm) by an average of 80\%.  MC methods increased both the RMS trajectory error and percentage of large errors slightly.  Fig. 3c shows histograms of trajectory errors in all projections across all acquired datasets (amplitudes ranging from 1 mm to 25 mm in 1 mm increments).  Trajectory error was less than 1mm in 92\% of projections (Phase and PhaseMC) and 82\% of projections (Base).

 \begin{figure*}
 \includegraphics{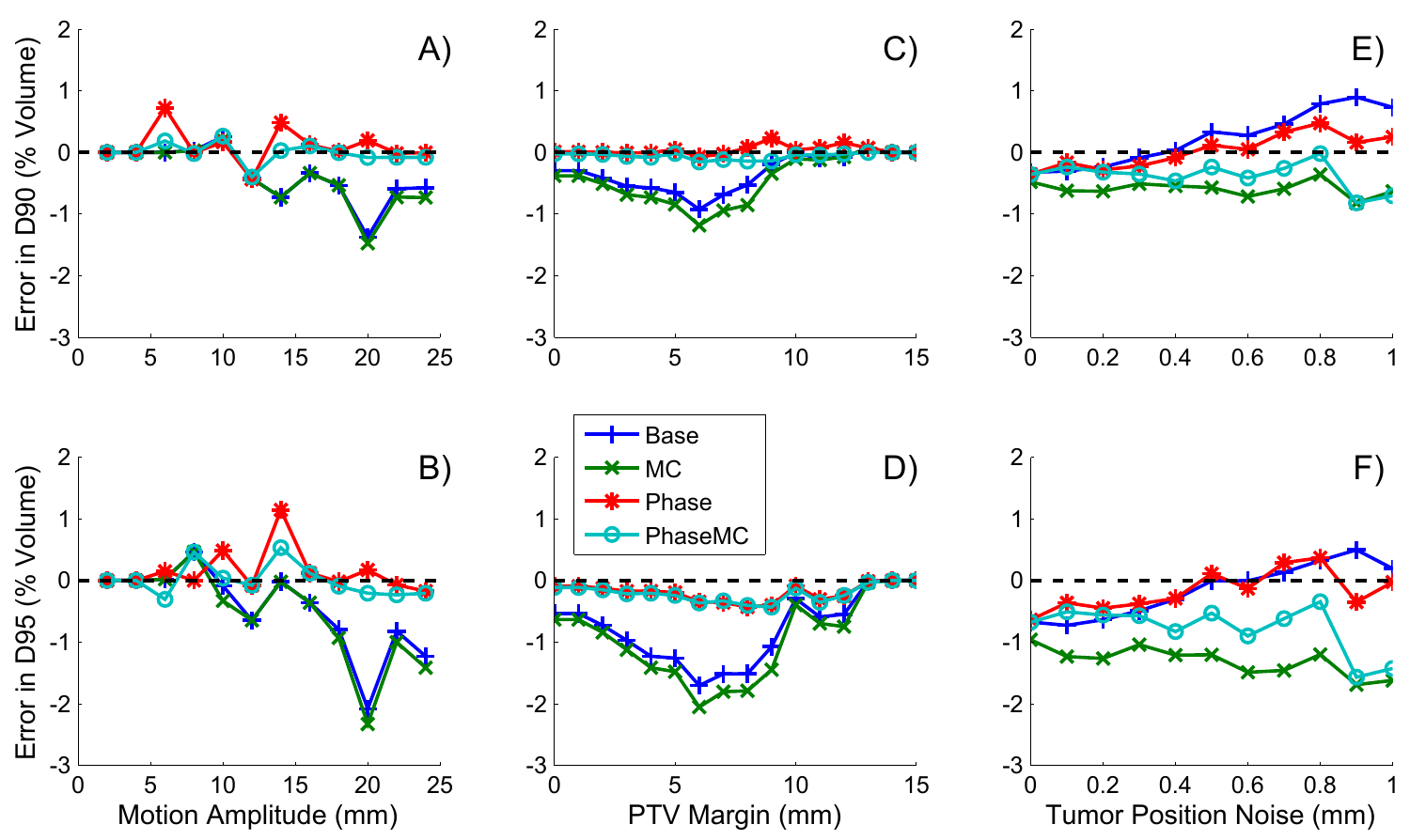}%
 \caption{Accuracy of calculating clinical dose metrics D90 and D95 from reconstructed trajectories.  Errors are calculated in 32 clinically-drawn CTVs. Motion amplitude ranged from 0 to 2.5 cm, and results were calculated with a range of isotropic margins and for trajectories with added noise.  Phase-resolved methods increased the accuracy of these dose metric calculations. As trajectory noise increased, the Base and Phase methods tended to overestimate D90 and D95, while MC methods prevented overestimation of dose.  In A) and B), the noise is 0 mm and the margin is 4 mm. In C) and D), the amplitude is 2.5 mm and the noise is 0 mm.  In E) and F), the amplitude is 1.2 mm and the margin is 4 mm.}
 \end{figure*}

\subsection{Dosimetric Accuracy}

Fig. 4 shows the distribution of mean dose errors and mean absolute dose errors using trajectories reconstructed with the Base, MC, Phase, and PhaseMC methods.  Results are shown across variations in trajectory amplitude, PTV margin size, and amount of added trajectory noise.  

Dose errors increased sharply as the motion amplitude increased, although in the Phase and PhaseMC methods, absolute error never exceeded 0.5\% of the prescription dose.  In general, dose errors decreased as the margin size increased, mainly due to an overall reduction in the contribution of motion seen by having more margin around the CTV.  However, it can also be seen that mean dose error was lower when using Phase and PhaseMC trajectory estimation methods.  This is likely due to the lower RMS trajectory error demonstrated by these methods.  
As gaussian trajectory noise increased, the expectation-based methods Base and Phase tended to overestimate dose, while Monte Carlo methods MC and PhaseMC tended to underestimate it. While Monte Carlo methods prevented the overestimation of dose, they also slightly increased the overall absolute error ($<$0.1\%).

Fig. 5 shows the accuracy of each method in determing dosimetric parameters D90 and D95 for varying amplitudes, PTV margin sizes, and trajectory noise.  One can see that the error of Phase methods was lower than the error in methods which ignore respiratory phase.  Expectation-based methods again overestimated the dose (relative to the true value) and Monte-Carlo methods underestimated it, but the Base and Phase methods had a lower absolute error.  The greatest accuracy was seen in the Phase method, which was able to determine D90 and D95 of the CTV to within ±0.5\% relative to prescription.  

\section{Discussion}

In this work, we have developed and tested methods to evaluate dose to a moving tumor using data from a standard clinical cone-beam CT.  In particular, we have tailored these methods to the problem of calculating dose to highly mobile tumors implanted with fiducial markers, such as lesions in the lung, pancreas, or liver\super{16-18}.  The trajectory and subsequent dose received by the target volume was determined with minimal error. Errors in mean position did not exceed 0.1 mm, and the RMS trajectory error averaged 3.8\% of the total range of motion.  Using the Phase method, the mean dose and mean absolute dose were calculated with over 99\% accuracy, and D90/D95 were determined to within 0.5\% of the true value.  

By incorporating information related to the respiratory cycle, the accuracy of tumor trajectory reconstruction and subsequent dosimetric analysis was improved.  The 3D Gaussian PDF utilized in this study stems from the work of Papiez and Langer\super{25},  where it was assumed that any systematic position errors are corrected with daily image guidance, and that only random fluctuations remain.  Thus the central limit theorem dictates that the distribution of the remaining random errors is Gaussian.  However, this assumption is broken for tumors which undergo regular respiratory motion, and the distribution of the trajectory about the mean position is no longer well definied as Gaussian.  By creating a unique Gaussian distribution for each portion of the respiratory cycle, this effect is minimized, and the accuracy of trajectory reconstruction is improved.  This method does have limitations; it cannot be used for tumors which undergo non-cyclical motion, such as prostate tumors.  It is also difficult to determine the respiratory phase in real-time.

Given the prevalence of CBCT, this method may provide a useful technique to clinically estimate the dose-of-the-day received by a mobile tumor, since over half of radiation oncology clinics perform some cone-beam CT imaging for localization\super{30}.  Our analysis is directed towards tumors implanted with fiducial markers, but also holds in cases without fiducials as long as the tumor location can be discerned in each projection (as has been demonstrated in lung tumors\super{31,32}).  In theory, our method allows for very accurate trajectory/dose reconstruction, since data are obtained using a much more comprehensive motion profile than other techniques.  A CBCT projection dataset constitutes upwards of 600 samples of tumor position, whereas 4D CT typically uses only 10 respiratory phases, and has been shown to not accurately represent daily intrafraction motion of the abdomen\super{33,34}.  As radiotherapy becomes more conformal, and hypofractionation becomes more commonplace, it becomes crucial to understand and account for the effects of tumor motion on dosimetry.  This method could be used to design patient-specific, optimized margins for treatment, or to verify the adequacy of the margins used.  It could also be used to decide between motion management strategies on a patient-specific basis (such as breath-hold, gating, or free-breathing).  It should be said that while we have validated the accuracy of our method in a simple clinical scenario, true dose-of-the-day reconstruction would require a more sophisticated dosimetric approach.  By constructing a simplified clinical scenario which maximizes the impact of tumor motion, our results indicate that tumor trajectories can be derived from CBCT projection datasets with sufficient accuracy to determine true dose. An additional benefit of this method is that the processing time is short.  Template-matching to determine seed locations can be performed in real-time, and it takes roughly 2 seconds to calculate the trajectory once the projected fiducial locations are known.  However, to build the 3D PDF, one must know the entire set of projected locations; in other words, the PDF cannot be built until after CBCT acquisition is completed.  Our method could be applied to generate an accurate estimate of tumor dosimetry shortly after the completion of the CBCT acquisition.

The conditions of this study were chosen to test the boundaries of dose reconstruction in highly mobile tumors.  Ranging up to 2.5 cm, the amplitude of motion in the SI direction corresponded to the displacement of very highly mobile lower lobe lung tumors, and exceeds the range of liver and other abdominal tumors\super{11,35}.  Hysteresis was also exaggerated, and motion in the AP and LR directions (up to 1.2 and 0.6 cm) exceeded that seen in clinical cases\super{11,35}.  In addition to the large range of motion, the effects of that motion on the dose were exaggerated by assuming perfect dose fall-off outside the PTV.  Our methods were also tested over a wide range of tumor volumes, and with up to 1 mm of Gaussian noise added to each point in the trajectory.  Even under these circumstances, dose metrics of the tumor were determined accurately.  One may note that the RMS trajectory error reported in this work (1-2 mm) exceeds the mean error in the similar work of Poulsen \textit{et al}\super{36}.  This can be explained by the large amplitude of motion in the current study, and agrees well with the maximum RMS error ($>$ 2 mm) reported for lung tumors in that work.  Additionally, these results give insight as to the scale of motion which causes the greatest dosimetric inaccuracy for each method.  In Fig. 4C, errors increase for the Base and MC methods as the magin size approaches the range of 5-8 mm, while the Phase and PhaseMC see the greatest error for 4-5 mm margins.  Since our dosimetric model is sensitive to motion equal to or greater than the PTV margin, and the frequency of motion errors decreases exponentially for larger errors (Fig 3C), these results suggest that motion on the order of 7 mm has the grestest effect for the Base and MC methods, and on the order of 4 mm for the Phase and PhaseMC methods.

It is clear from these results that incorporation of respiratory phase information is beneficial when reconstructing the tumor trajectory of highly mobile tumors.  However, the role of MC sampling is mixed.  In the Base scenario, the position of the tumor is determined as the expectation value of the Gaussian PDF along the detection ray.  Yet one expects this determination to exhibit a Gaussian error profile, as the tumor position undergoes random fluctuations in addition to the cyclic respiratory motion.  MC sampling addresses this by sampling a greater portion of the tails of this distribution (relative to using the expectation value), thereby generating a distribution of positions with a more realistic proportion of large deviations.  One may expect the MC method to increase trajectory error; however, under certain conditions one would expect a more complete sampling of the error distribution to result in a more accurate calculation of derived metrics (such as target dosimetry).  As the trajectory noise increases, the Base and Phase methods to overestimate the true tumor dose, as the expectation value of the distribution likely lies within the PTV.  MC sampling slightly increases error in an absolute sense (Fig. 4), but is a conservative estimate of D90/D95 target dose metrics in noisy trajectories (Fig. 5).  MC methods may be useful for calculating dose to the prescription volume, and could constitute a worst-case scenario estimate of tumor dose.  However, for organs-at-risk and other volumes, one would likely use expectation-based position estimation (i.e. Base or Phase), which yields a more accurate result.

\section{Conclusions}

We have developed methods for accurately determining the trajectory of highly-mobile abdominal tumors using cone-beam CT projections.  These methods have been applied to reconstruct the 3D trajectory of a respiratory motion phantom, and to calculate the dose received under that trajectory.  Results were calculated without respiratory phase or Monte Carlo corrections (Base), with phase corrections only (Phase), with Monte Carlo corrections only (MC), and with both phase and Monte Carlo corrections (PhaseMC).  In all methods, errors in mean position did not exceed 0.1 mm, and the RMS trajectory error was only 3.8\% of the total range of motion.  Using phase-resolved methods Phase and PhaseMC, the mean dose and mean absolute dose were calculated with over 99\% accuracy, and D90/D95 were determined to within 0.5\% relative to the prescription dose.  These methods can be applied to estimate the dose-of-the-day for mobile tumors, and can aid in the selection of motion management strategies.

\section{Acknowledgements}
The authors would like to thank Hirotatsu Armstrong for several helpful discussions and comments regarding this work.

\section*{References}
{\small 
1.	P. J. Keall, G. S. Mageras, J. M. Balter, R. S. Emery, K. M. Forster, S. B. Jiang, J. M. Kapatoes, D. A. Low, M. J. Murphy and B. R. Murray, "The management of respiratory motion in radiation oncology report of AAPM Task Group 76," Medical Physics 33, 3874 (2006).

2.	S. H. Benedict, K. M. Yenice, D. Followill, J. M. Galvin, W. Hinson, B. Kavanagh, P. Keall, M. Lovelock, S. Meeks and L. Papiez, "Stereotactic body radiation therapy: the report of AAPM Task Group 101," Medical Physics 37, 4078 (2010).

3.	D. A. Jaffray, J. H. Siewerdsen, J. W. Wong and A. A. Martinez, "Flat-panel cone-beam computed tomography for image-guided radiation therapy," Int J Rad Oncol* Biol* Phys 53, 1337-1349 (2002).

4.	T. Pan, T.-Y. Lee, E. Rietzel and G. T. Chen, "4D-CT imaging of a volume influenced by respiratory motion on multi-slice CT," Medical Physics 31, 333 (2004).

5.	H. H. Liu, P. Balter, T. Tutt, B. Choi, J. Zhang, C. Wang, M. Chi, D. Luo, T. Pan and S. Hunjan, "Assessing respiration-induced tumor motion and internal target volume using four-dimensional computed tomography for radiotherapy of lung cancer," International Journal of Radiation Oncology* Biology* Physics 68, 531-540 (2007).

6.	C. R. Ramsey, D. Scaperoth, D. Arwood and A. L. Oliver, "Clinical efficacy of respiratory gated conformal radiation therapy," Medical Dosimetry 24, 115-119 (1999).

7.	P. Kupelian, T. Willoughby, A. Mahadevan, T. Djemil, G. Weinstein, S. Jani, C. Enke, T. Solberg, N. Flores and D. Liu, "Multi-institutional clinical experience with the Calypso System in localization and continuous, real-time monitoring of the prostate gland during external radiotherapy," International Journal of Radiation Oncology* Biology* Physics 67, 1088-1098 (2007).

8.	M. J. Murphy, "Tracking moving organs in real time," Seminars in Radiation Oncology 14, 91-100 (2004).

9.	P. J. Keall, H. Cattell, D. Pokhrel, S. Dieterich, K. H. Wong, M. J. Murphy, S. S. Vedam, K. Wijesooriya and R. Mohan, "Geometric accuracy of a real-time target tracking system with dynamic multileaf collimator tracking system," International Journal of Radiation Oncology* Biology* Physics 65, 1579-1584 (2006).

10.	H. Shirato, K. Suzuki, G. C. Sharp, K. Fujita, R. Onimaru, M. Fujino, N. Kato, Y. Osaka, R. Kinoshita and H. Taguchi, "Speed and amplitude of lung tumor motion precisely detected in four-dimensional setup and in real-time tumor-tracking radiotherapy," International Journal of Radiation Oncology* Biology* Physics 64, 1229-1236 (2006).

11.	Y. Seppenwoolde, H. Shirato, K. Kitamura, S. Shimizu, M. van Herk, J. V. Lebesque and K. Miyasaka, "Precise and real-time measurement of 3D tumor motion in lung due to breathing and heartbeat, measured during radiotherapy," International Journal of Radiation Oncology* Biology* Physics 53, 822-834 (2002).

12.	B. Cho, P. R. Poulsen, A. Sloutsky, A. Sawant and P. J. Keall, "First demonstration of combined kV/MV image-guided real-time dynamic multileaf-collimator target tracking," International Journal of Radiation Oncology* Biology* Physics 74, 859-867 (2009).

13.	K. Malinowski, T. J. McAvoy, R. George, S. Dietrich and W. D. D’Souza, "Incidence of changes in respiration-induced tumor motion and its relationship with respiratory surrogates during individual treatment fractions," International Journal of Radiation Oncology* Biology* Physics 82, 1665-1673 (2012).

14.	A. P. Shah, P. A. Kupelian, B. J. Waghorn, T. R. Willoughby, J. M. Rineer, R. R. Mañon, M. A. Vollenweider and S. L. Meeks, "Real-Time Tumor Tracking in the Lung Using an Electromagnetic Tracking System," International Journal of Radiation Oncology* Biology* Physics (2013).

15.	J. Yun, K. Wachowicz, M. Mackenzie, S. Rathee, D. Robinson and B. Fallone, "First demonstration of intrafractional tumor-tracked irradiation using 2D phantom MR images on a prototype linac-MR," Medical Physics 40, 051718 (2013).

16.	W. G. Park, B. M. Yan, D. Schellenberg, J. Kim, D. T. Chang, A. Koong, C. Patalano and J. Van Dam, "EUS-guided gold fiducial insertion for image-guided radiation therapy of pancreatic cancer: 50 successful cases without fluoroscopy," Gastrointestinal endoscopy 71, 513-518 (2010).

17.	W. Wunderink, A. Méndez Romero, W. De Kruijf, H. De Boer, P. Levendag and B. Heijmen, "Reduction of respiratory liver tumor motion by abdominal compression in stereotactic body frame, analyzed by tracking fiducial markers implanted in liver," International Journal of Radiation Oncology* Biology* Physics 71, 907-915 (2008).

18.	M. Imura, K. Yamazaki, H. Shirato, R. Onimaru, M. Fujino, S. Shimizu, T. Harada, S. Ogura, H. Dosaka-Akita and K. Miyasaka, "Insertion and fixation of fiducial markers for setup and tracking of lung tumors in radiotherapy," International Journal of Radiation Oncology* Biology* Physics 63, 1442-1447 (2005).

19.	C. Wu, R. Jeraj, G. H. Olivera and T. R. Mackie, "Re-optimization in adaptive radiotherapy," Physics in Medicine and Biology 47, 3181 (2002).

20.	G. D. Hugo, D. Yan and J. Liang, "Population and patient-specific target margins for 4D adaptive radiotherapy to account for intra-and inter-fraction variation in lung tumour position," Physics in Medicine and Biology 52, 257 (2007).

21.	C. Ozhasoglu and M. J. Murphy, "Issues in respiratory motion compensation during external-beam radiotherapy," International Journal of Radiation Oncology* Biology* Physics 52, 1389-1399 (2002).

22.	P. R. Poulsen, B. Cho and P. J. Keall, "Real-time prostate trajectory estimation with a single imager in arc radiotherapy: A simulation study," Physics in medicine and biology 54, 4019 (2009).

23.	P. R. Poulsen, B. Cho, K. Langen, P. Kupelian and P. J. Keall, "Three-dimensional prostate position estimation with a single x-ray imager utilizing the spatial probability density," Physics in medicine and biology 53, 4331 (2008).

24.	R. Li, B. P. Fahimian and L. Xing, "A Bayesian approach to real-time 3D tumor localization via monoscopic x-ray imaging during treatment delivery," Medical physics 38, 4205 (2011).

25.	L. Papiez and M. Langer, "On probabilistically defined margins in radiation therapy," Physics in Medicine and Biology 51, 3921 (2006).

26.	L. A. Feldkamp, L. C. Davis and J. W. Kress, "Practical cone-beam algorithm," J Opt Sci Am A 1, 612-619 (1984).

27.	J. C. Lagarias, J. A. Reeds, M. H. Wright and P. E. Wright, "Convergence properties of the Nelder--Mead simplex method in low dimensions," SIAM Journal on optimization 9, 112-147 (1998).

28.	J. Lewis, "Fast template matching," Vision interface 95, 15-19 (1995).

29.	J. Lewis, "Fast normalized cross-correlation," Vision interface 10, 120-123 (1995).

30.	D. R. Simpson, J. D. Lawson, S. K. Nath, B. S. Rose, A. J. Mundt and L. K. Mell, "A survey of image‐guided radiation therapy use in the United States," Cancer 116, 3953-3960 (2010).

31.	A. Richter, J. Wilbert, K. Baier, M. Flentje and M. Guckenberger, "Feasibility study for markerless tracking of lung tumors in stereotactic body radiotherapy," International Journal of Radiation Oncology* Biology* Physics 78, 618-627 (2010).

32.	J. Rottmann, M. Aristophanous, A. Chen and R. Berbeco, "A multi-region algorithm for markerless beam's-eye view lung tumor tracking," Physics in medicine and biology 55, 5585 (2010).

33.	J. Ge, L. Santanam, C. Noel and P. J. Parikh, "Planning 4-Dimensional Computed Tomography (4DCT) Cannot Adequately Represent Daily Intrafractional Motion of Abdominal Tumors," Int J Rad Oncol* Biol* Phys (2012).

34.	A. Y. Minn, D. Schellenberg, P. Maxim, Y. Suh, S. McKenna, B. Cox, S. Dieterich, L. Xing, E. Graves and K. A. Goodman, "Pancreatic tumor motion on a single planning 4D-CT does not correlate with intrafraction tumor motion during treatment," American Journal of Clinical Oncology 32, 364 (2009).

35.	H. Shirato, Y. Seppenwoolde, K. Kitamura, R. Onimura and S. Shimizu, "Intrafractional tumor motion: lung and liver," Seminars in radiation oncology 14, 10-18 (2004).

36.	P. R. Poulsen, B. Cho and P. J. Keall, "A method to estimate mean position, motion magnitude, motion correlation, and trajectory of a tumor from cone-beam CT projections for image-guided radiotherapy," International Journal of Radiation Oncology* Biology* Physics 72, 1587-1596 (2008).
}


\end{document}